\documentclass[aps,twocolumn,nofootinbib,superscriptaddress]{revtex4}

\usepackage{slashed}
\usepackage{epsfig,latexsym,cancel,amssymb,amsmath,verbatim,mathrsfs}
\usepackage{color,xcolor,graphicx}
\usepackage{multirow}
\usepackage{diagbox}
\usepackage{hyperref}
\hypersetup{
	colorlinks,
	linkcolor={red!70!black},
	citecolor={blue!80!black},
	urlcolor={blue!80!black},
	bookmarksopen=true
}
\newcommand{\beq}{\begin{equation}}
\newcommand{\eeq}{\end{equation}}
\newcommand{\bea}{\begin{eqnarray}}
\newcommand{\eea}{\end{eqnarray}}
\newcommand{\met}{\slashed{\rm E}_{\rm T}}
\newcommand{\nn}{\nonumber}

\definecolor{darkgreen}{HTML}{228B22}

\allowdisplaybreaks[4]

\begin{document}

\title{New Physics and Two Boosted $W$-jets plus Missing Energy}

\author{Qing-Hong Cao}
\email{qinghongcao@pku.edu.cn}
\affiliation{School of Physics and State Key Laboratory 
of Nuclear Physics and Technology, Peking University, Beijing 100871, China}
\affiliation{Collaborative Innovation Center of Quantum Matter, Beijing 100871, China}
\affiliation{Center for High Energy Physics, Peking University, Beijing 100871, China}

\author{Nuo Chen}
\email{1701110064@pku.edu.cn}
\affiliation{School of Physics and State Key Laboratory 
of Nuclear Physics and Technology, Peking University, Beijing 100871, China}

\author{Hao-Ran Jiang}
\email{h.r.jiang@pku.edu.cn}
\affiliation{School of Physics and State Key Laboratory 
of Nuclear Physics and Technology, Peking University, Beijing 100871, China}

\author{Bin Li}
\email{libin@pku.edu.cn}
\affiliation{School of Physics and State Key Laboratory 
of Nuclear Physics and Technology, Peking University, Beijing 100871, China}

\author{Yandong Liu}
\email{ydliu@bnu.edu.cn}
\affiliation{Key Laboratory of Beam Technology of Ministry of Education, College of Nuclear Science and Technology, Beijing Normal University, Beijing 100875, China}
\affiliation{Beijing Radiation Center, Beijing 100875, China}

\begin{abstract}
We show that the signature of two boosted $W$-jets plus large missing energy is very promising  to probe heavy charged resonances ($X^\pm$) through the process of $pp\to X^+X^-\to W^+W^- X^0 X^0$ where $X^0$ denotes dark matter candidate. The hadronic decay mode of the $W$ boson is considered to maximize the number of signal events. When the mass split between $X^\pm$ and $X^0$ is large, one has to utilize the jet-substructure technique to analyze the boosted $W$-jet. For illustration we consider the process of chargino pair production at the LHC, i.e., $pp\to \chi_1^+\chi^-_1 \to W^+W^-\chi_1^0\chi_1^0$, and demonstrate that the proposed signature is able to cover more parameter space of $m_{\chi_1^\pm}$ and $m_{\chi_1^0}$ than the conventional signature of multiple leptons plus missing energy.  More importantly, the signature of our interests is not sensitive to the spin of heavy resonances. 
\end{abstract}

\maketitle

\noindent {\bf 1. Introduction.}

The weakly interacting massive particle (WIMP) is one of the promising candidate of dark matter (DM) whose existence has been confirmed from various gravity effects~\cite{Rubin:1970zza,Rubin:1980zd,Aghanim:2018eyx}. Until now only null results are reported by DM searching experiments which impose stringent bounds on the DM candidate~\cite{Aprile:2017aty,Akerib:2016vxi,Wang:2020coa,Zyla:2020zbs}. 
One way to probe the DM candidate ($X^0$) at the Large Hadron Collider (LHC) is through the so-called ``mono-X" channel in which a pair of DM candidates is produced in association with a jet or a photon radiated out from initial state partons~\cite{Aaboud:2017dor,Sirunyan:2017ewk,Aad:2020arf,Aaboud:2017phn,Sirunyan:2017jix,Aad:2021egl,Aaboud:2017bja,Sirunyan:2017hci,Cao:2018nbr,Cao:2020ihv,Sirunyan:2020fwm,Sirunyan:2019zav,Aad:2020sef}. However, as strongly correlated with DM direct detection experiments, the mono-X channels are highly constrained in DM direct search experiments~\cite{Cao:2009uw,Goodman:2010ku}. Other ways are through the process of  $pp\to X^\pm X^0$ and $pp\to X^+X^-$, where $X^\pm$ denotes the next-to-lightest dark particle. The collider signature relies on the mass split $\Delta m\equiv m_{X^\pm}-m_{X^0}$. If $\Delta m$ is much less than electroweak (EW) gauge bosons, i.e., $\Delta m\ll m_{W/Z}$, it yields a signal of long-lived particles or suddenly disappearing tracks~\cite{Sirunyan:2020pjd,Belyaev:2020wok}. When $\Delta m > m_{W/Z}$, it yields a collider signature of missing transverse momentum ($\met$) plus either  leptons
~\cite{Aaboud:2018sua,Aad:2019vvi,Aad:2019vnb,Sirunyan:2018lul} 
or jets~\cite{Sirunyan:2017qaj,CMS:2020wxd}, depending on how the EW gauge boson decays, where $\met$ predominantly  originates from the DM candidates. Suffering from huge SM backgrounds, the potential of the process of $pp\to X^\pm X^0$  on probing the DM is limited~\cite{Bhardwaj:2019mts}.

\begin{figure}
\includegraphics[scale=0.5]{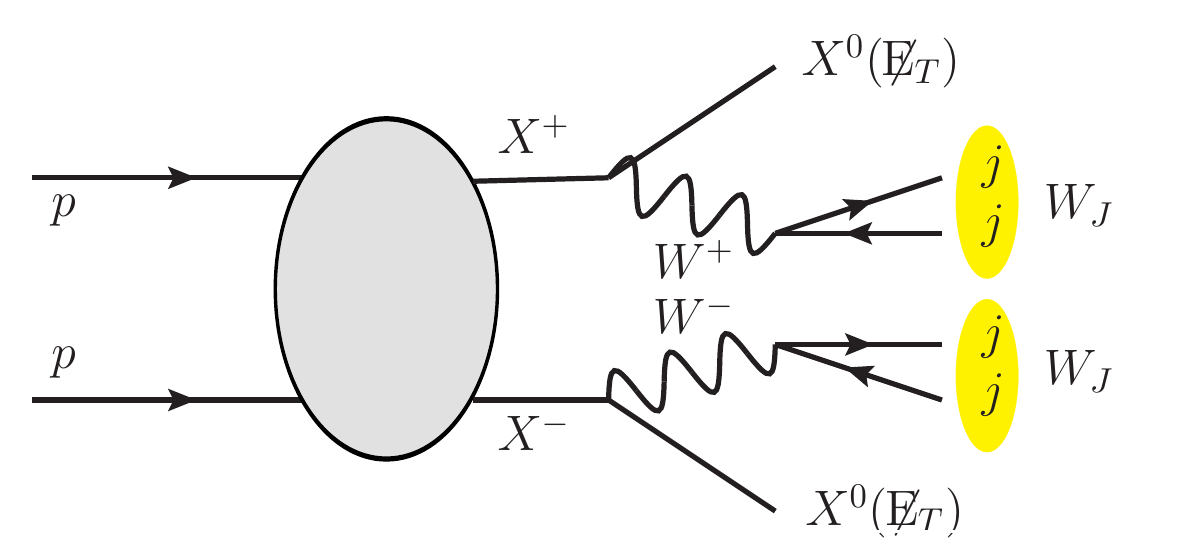}
\caption{ Pictorial Illustration of the two boosted $W$-jet plus missing energy $\met$ production at the LHC.}
\label{fig:illustrative}
\end{figure}

We consider the process of $pp\to X^+X^-$ with subsequent decays of $X^\pm\to W^\pm X^0$; see Fig.~\ref{fig:illustrative}. In order to increase the signal rate, we demand both the $W$ bosons decaying into a pair of quarks. When the mass split $\Delta m$ is much larger than $m_{W}$, the $W$ boson is highly boosted such that the quarks from its decay are collimated and form a fat $W$-jet ($W_J$). The conventional jet reconstruction method no longer works and one has to adopt the jet-substructure method to deal with the fat $W$-jet. When the transverse momentum of $W$ is larger than 200 GeV, the jet-substructure algorithm dominates over the traditional method in the aspects of identifying the $W$-jet and suppressing QCD backgrounds~\cite{Khachatryan:2014vla}. 
Hence, we focus on the collider signature of two fat $W$-jets plus $\met$, denoted as $W_JW_J\met$, and show that it works much better than the conventional method in probing $X^0$ and $X^\pm$, especially when the mass split $\Delta m$ is large.  

The dark particles can be  scalars, fermions or vectors which arise from various new physics models; they can be the dark scalar $S^\pm $ in the Inert Doublet Higgs models, the fermionic supersymmetric particle $F^\pm$ in the SUSY models~\cite{Ma:2006km,Barbieri:2006dq,Cao:2007rm}, or the additional gauge boson $V^\pm$ in the Little Higgs models~\cite{ArkaniHamed:2008qn,Goudelis:2013uca,Belanger:2015kga}.
Our study shows that the signature of $W_JW_J\met$ is not sensitive to the spin of the DM candidate and thus can be extensively used in the DM searches. 

~\\
\noindent{ \bf 2. Collider simulation. }

\begin{figure*}
\centering
\includegraphics[scale=.8]{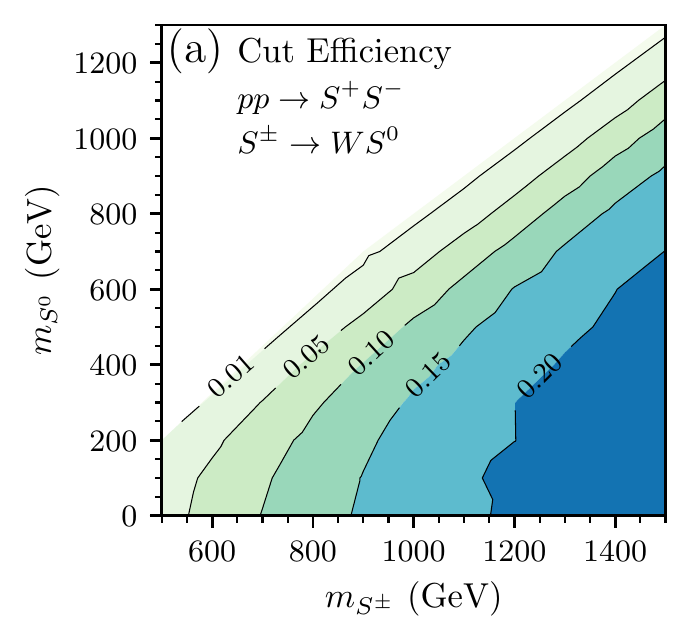}\includegraphics[scale=.8]{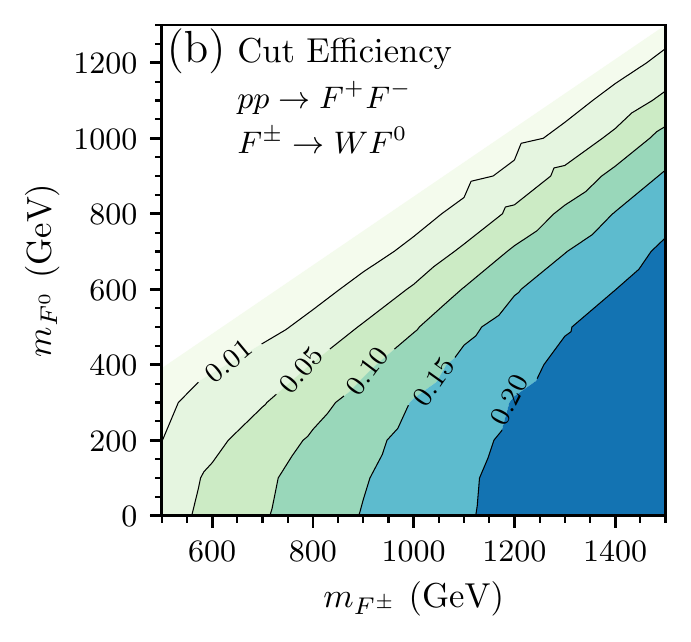}\includegraphics[scale=.8]{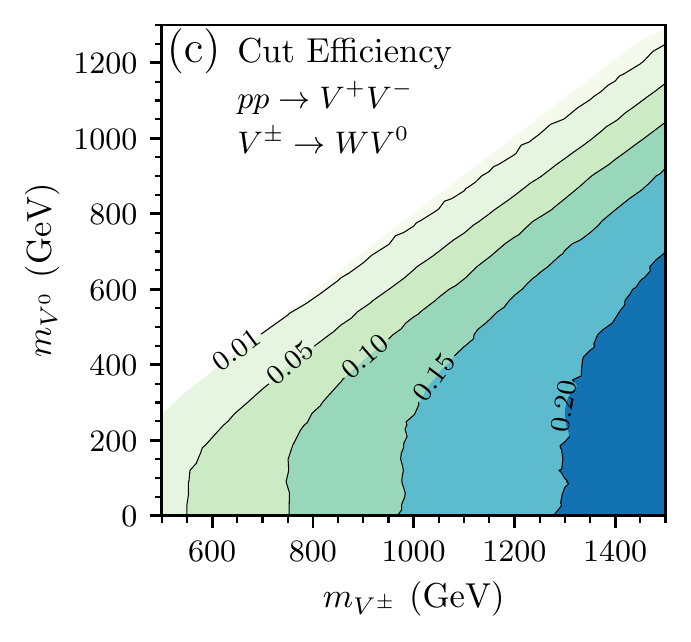}
\caption{Efficiencies after applying all the cuts in the production of the scalar pairs $S^+S^-$ (a), the fermion pairs $F^+F^-$ (b) and the vector pairs $V^+V^-$ (c). }
\label{fig:cut-efficiency}
\end{figure*}

We perform a collider simulation at the LHC with a collider energy of 14~TeV. 
For simplicity we assume that the $X^\pm$ decays entirely into a pair of $X^0$ and $W^\pm$ and demand the $W$ boson decaying into quarks to enlarge the branching ratio. The signal and background events are generated using MadGraph5 \cite{Alwall:2014hca} and then linked with Pythia~\cite{Sjostrand:2006za} and Delphes~\cite{deFavereau:2013fsa} for parton shower, hadronization and detector simulation. In order to trigger the signal event, we demand exactly two $W$-jets and no leptons in each event, i.e., 
\begin{equation}
\label{equ:num}
N^{\ell}=0,~~~N^{J}=2.
\end{equation}
The $W$-jet is reconstructed using  FastJet~\cite{Cacciari:2011ma,Cacciari:2005hq} in terms of the anti-$k_t$ algorithm~\cite{Cacciari:2008gp} with the radius $R=0.8$. The N-subjettiness algorithm~\cite{Thaler:2010tr} is used to suppress QCD backgrounds. In addition the trimming~\cite{Krohn:2009th}, Pruning~\cite{Ellis:2009su} and SoftDrop~\cite{Larkoski:2014wba} techniques are utilized to further groom away soft radiation in the resulting fat jets and polish the mass of jets. 

The invariant masses ($m_J$) of reconstructed 2-pronged $W$-jets are required to be within the mass window of~\cite{Cao:2015cdb}
\begin{equation}\label{equ:mw}
m_W-13~\text{GeV} \leq m_J \leq m_W + 13~\text{GeV}.
\end{equation}
We order the two $W$-jets by their transverse momentum ($p_T$) as the leading fat-jet $J_1$ and the next-to-leading fat-jet $J_2$. In order to 
ensure that the $W$-jets are indeed boosted and tagged in the central region of the detector, we further impose large $p_T$ cuts on the two $W$-jets as follows:
\begin{equation}\label{equ:pt}
p_T^{J_1, J_2}\ge 200~ \textrm{GeV},~~~|\eta^{J_1,J_2}|\le3, 
\end{equation}
where $\eta^{i}$ denotes the rapidity of the $i$-th jet. 
Finally, we impose hard cuts on the $\met$ and invariant mass of the two reconstructed $W$-jets, 
\begin{equation}
\label{equ:met}
\met \ge 400~\text{GeV},~~~m_{J_1,J_2}\ge500~\text{GeV}.
\end{equation}
to help extracting the signal out of the SM backgrounds.

The signature of our interests is not sensitive to the spin of the $X^\pm$ particle. 
Figure~\ref{fig:cut-efficiency} shows the efficiency of the signal event surviving all the cuts in the plane of $m_{X^\pm}$ and $m_{X^0}$. It is clear that the contour lines are approximately linear in the region of $200~{\rm GeV} \lesssim  \Delta m \lesssim 800~{\rm GeV}$ as the reconstruction efficiency mainly depends on the mass splitting $\Delta m$. In addition the efficiency contour is less sensitive to the dark matter mass when $m_{X^0}\leq 200~{\rm GeV}$, for all the cases of three different spins.

The SM backgrounds predominantly arise from four sources as follows: 1) the pair production of $WW$, $WZ$ and $ZZ$ bosons; 2) the $t\bar{t}$ production; 3) the associated production of a $W$ boson with multiple jets (denoted by $W$+jets); 4) the associated production of a $Z$ boson with multiple jets ($Z$+jets). The backgrounds from $W$+jets, $Z$+jets and triple gauge boson production are negligible after kinematic cuts. The numbers of the signal and background events at the 14~TeV LHC with an integrated luminosity ($\mathcal{L}$) of 100 fb$^{-1}$ before and after the event reconstruction are listed as follows:
\begin{align}
\hline
~  &&    t\bar{t} && WW && WZ && ZZ  \nn\\ 
{\rm Before} && 5.5\times10^7 && 9.54\times10^6 && 4.36\times10^6 && 1.25\times10^6 \nn\\
{\rm After}  && 96.25 && 19.35 && 50.0 && 6.57 \nn\\
\hline\nn
\end{align}
The $t\bar{t}$ production dominates owing to its large rate at the LHC. 
Given the numbers of the background event and the cut efficiencies of the signal event, we are ready to obtain the upper limit of the signal event at the $95\%$ confidence level in terms of 
\begin{equation}
\sqrt{-2 \left(n_b \ln \frac{n_s + n_b}{n_b}-n_s \right)} = 2.0,
\label{eq:significance}
\end{equation}
where $n_s$ and $n_b$ is the number of signal and background event, respectively.

\noindent {\bf 3. A demo in SUSY search.} 

To demonstrate the power of our method, we explore the potential of the LHC to search for the production of chargino ($\chi_1^\pm$) pair in a simplified supersymmetric extension of the SM (SUSY) in which the $\chi_1^\pm$ is assumed to entirely decay via the mode of $\chi_1^\pm \to W^\pm + \chi_1^0$. The neutralino $\chi_1^0$ is  the DM candidate. 

\begin{figure}
\includegraphics[scale=0.58]{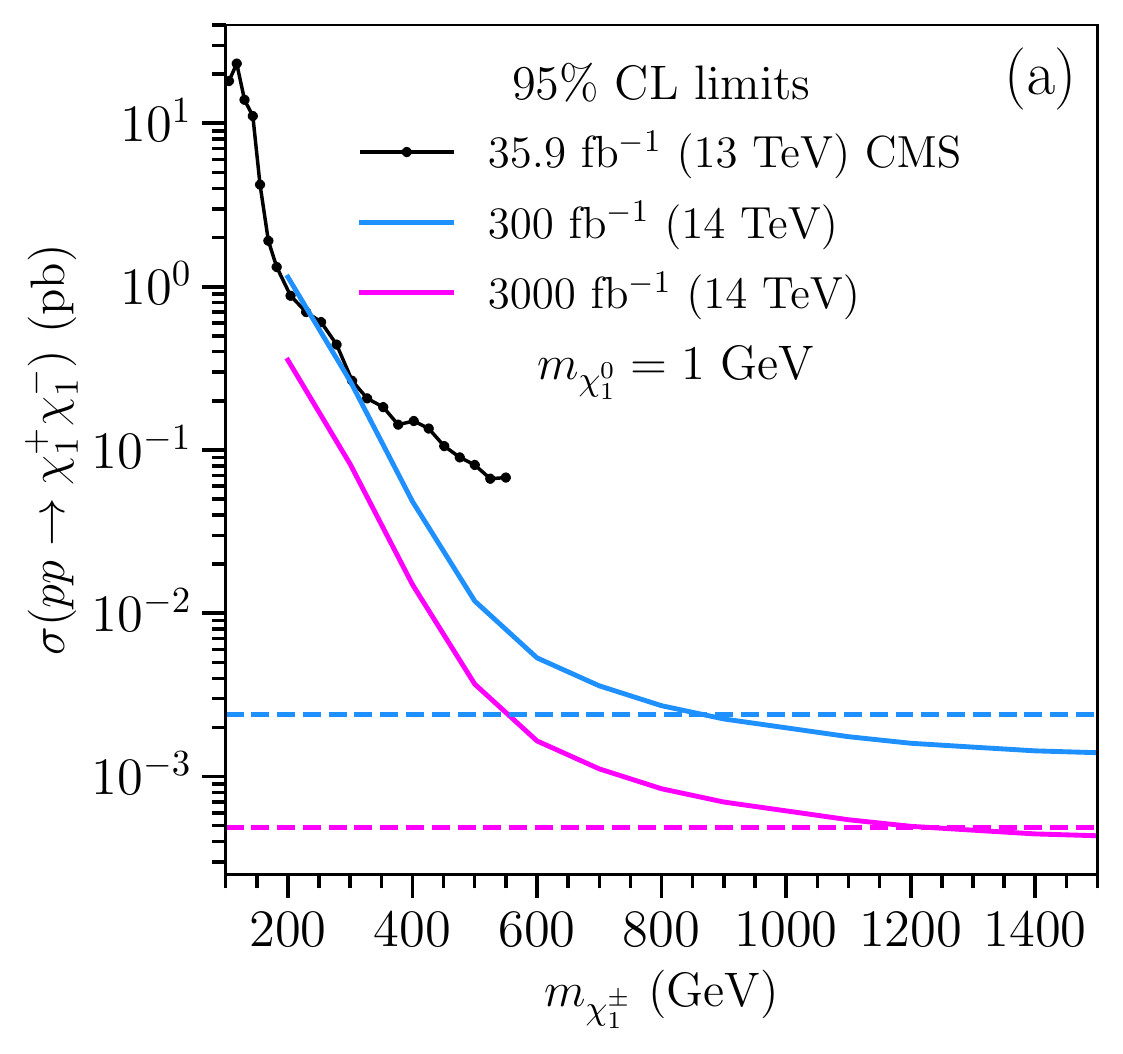}
\includegraphics[scale=0.58]{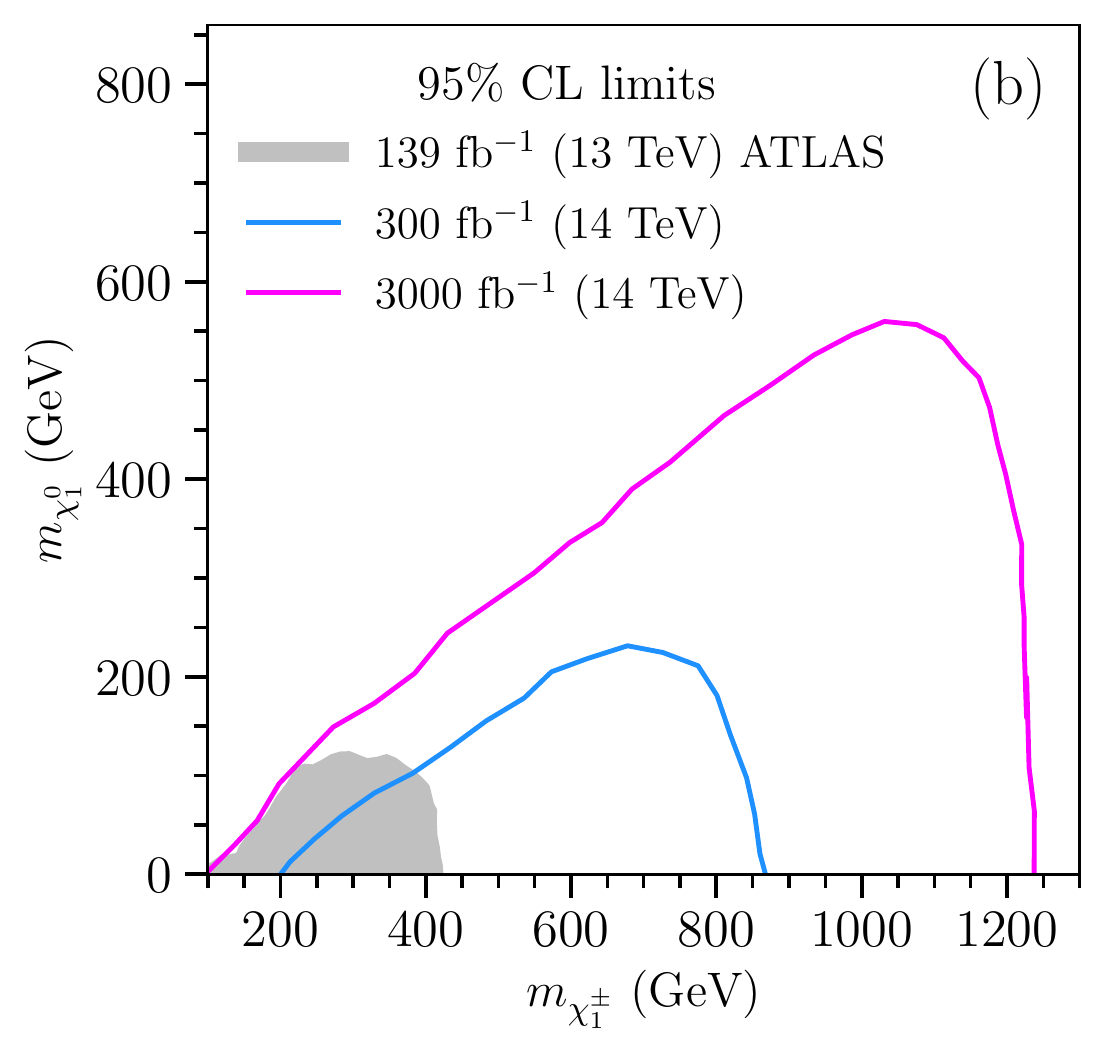}
\caption{(a) Upper limits of $\sigma(\chi_1^+ \chi_1^-)$ production as a function of $m_{\chi_1^\pm}$ at the 95\% confidence level at the LHC with an integrated luminosity of $300~{\rm fb}^{-1}$ (blue) and $3000~{\rm fb}^{-1}$ (magenta) with $m_{\chi_1^0}=0$ and ${\rm Br}(\chi_1^+\to W^+\chi_1^0)=1$. The black dotted curve denotes the current limit obtained from the mode of multiple lepton plus $\met$. (b) Contours of the upper limit of  $\sigma(\chi_1^+\chi_1^-)$ in the plane of $m_{\chi_1^\pm }$ and $m_{\chi_1^0}$. }
\label{fig:xsection}
\end{figure}

When the mass split between the chargino and neutralino is sizable, say $\Delta m \gtrsim 100~\text{GeV}$, it yields a signature of multiple leptons plus $\met$ which has been measured by the ATLAS~\cite{Aaboud:2018sua,Aad:2019vvi,Aad:2019vnb} and the CMS collaboration~\cite{Sirunyan:2017qaj,Sirunyan:2018lul}. 
In the analysis of the CMS collaboration, the neutralino mass is set to be 1 GeV, i.e., $m_{\chi_1^0}=1$ GeV  in order to obtain an upper bound of $\sigma(\chi_1^+\chi_1^-)$ as a function of $m_{\chi^\pm_1}$ at the 13 TeV LHC with $\mathcal{L}=35.9~{\rm fb}^{-1}$~\cite{Sirunyan:2018lul}; see the black-dotted curve in Fig.~\ref{fig:xsection}(a). 
Following the CMS group we set $m_{\chi_1^0}=1$ GeV and perform a collider simulation of the signature of the two boosted $W$-jets and $\met$. The blue and magenta curves denote the projected upper limit derived from the signature of two $W$-jets plus $\met$ at the 14~TeV LHC with $\mathcal{L}=300~{\rm fb}^{-1}$ and $3000~{\rm fb}^{-1}$, respectively. Obviously, the boosted $W$-jet method efficiently increases the sensitivity to the search of the $\chi_1^+\chi^-$ pair when $m_{\chi^\pm_1}\ge 200~{\rm GeV}$ for two reasons. One is the good efficiency of reconstructing the boosted $W$-jets, the other is the large suppression of the QCD backgrounds~\cite{Aad:2015rpa,Khachatryan:2016zcu,Aaboud:2017eta,Aaboud:2017fgj}. From Eq.~\ref{eq:significance} and the event numbers of the backgrounds given above, we obtain the $2\sigma$ bound on the $\sigma(\chi_1^+\chi_1^-)$ and $m_{\chi_1^\pm}$ for a massless $\chi_1^0$ as following:
\begin{align}
& \sigma(\chi_1^+\chi_1^-)\leq 2.4~{\rm fb}, &&m_{\chi_1^\pm}> 870~{\rm GeV},\nn
\end{align}
with $ \mathcal{L}=300~{\rm fb}^{-1}$ and 
\begin{align}
&\sigma(\chi_1^+\chi_1^-)\leq 0.48~{\rm fb}, &&m_{\chi_1^\pm}> 1240~{\rm GeV},\nn
\end{align}
with $\mathcal{L}=3000~{\rm fb}^{-1}$; see the blue-dashed and magenta-dashed horizontal lines.

The results of a massive $\chi^0_1$ are shown in Fig.~\ref{fig:xsection}(b) which plots the excluded regions in the plane of $m_{\chi_1^\pm}$ and $m_{\chi_1^0}$ at the $95\%$ confidence level.  The gray region is excluded in the signal signature of multi-leptons plus $\met$ at the 13 TeV LHC with an integrated luminosity of 139 fb$^{-1}$~\cite{Aad:2019vnb}. The region under the blue curve is excluded at the 14 TeV LHC with an integrated luminosity of 300~fb$^{-1}$. The efficiency of the $W$-jet reconstruction increases with $\Delta m=m_{\chi_1^\pm}-m_{\chi_1^0}$, and the production rate of $\chi_1^+\chi_1^-$ pair decreases rapidly with $m_{\chi_1^+}$. The two effects compete with each other and yield a peak around $m_{\chi_1^+}\sim 700~{\rm GeV}$; see the blue curve. Accumulating 10 times more data, i.e., increasing the integrated luminosity to $3000~{\rm fb}^{-1}$, the peak position is shifted to $m_{\chi_1^\pm}\sim 1100~{\rm GeV}$. It is evident that the signature of two $W$-jets plus $\met$ works much better than the conventional mode of multiple leptons plus $\met$; the magenta curve covers a vast parameter space consisting of the gray region. We thus urge that one should utilize the signature of the two $W$-jets plus $\met$ to search for a heavy resonance.

~\\
\noindent {\bf 4. Summary.} 

In this work we explore the potential of searching for the dark matter candidate $X^0$ through the pair production of heavy charged resonance $X^\pm$ which predominantly decays into a pair of $W^\pm$ and $X^0$, i.e., $pp\to X^+X^-\to W^+W^-X^0X^0$.  In order to achieve more signal events, we demand the $W$ boson decaying into a pair of quark. When the mass split between $X^\pm$ and $X^0$ is large, say $\Delta m=m_{X^\pm} - m_{X^0}\gg m_W$, the $W$ boson is boosted such that  the jet-substructure method is needed to increase the efficiency of the even reconstruction. The collider signature of our interests is two boosted $W$-jets plus $\met$ which originates from the two invisible dark matter candidates. We demonstrate that the signature is not sensitive to the spin of heavy charged resonance $X^\pm$ and the efficiency of event reconstruction mainly depends on $\Delta m$. Our method can be used in the search of various new physics resonances as long as they decay into a pair of $W$ bosons and the dark matter candidate.

For illustration we consider the process of chargino pair production in a simplified supersymmetric extension of the SM, i.e., $pp\to \chi_1^+\chi_1^-\to \chi_1^0\chi_1^0 W^+ W^-$. Our collider simulation shows that the signature of $W_JW_J\met$ works much better than the conventional mode of multiple leptons plus $\met$ at the LHC, especially in the region of $200~{\rm GeV}\leq m_{\chi^\pm_1} -m_{\chi_1^0 }\leq 800~{\rm GeV}$.  

\noindent{\bf Acknowledgements.}
The work is supported in part by the National Science Foundation of China under Grant Nos. 11725520, 11675002, 11635001, 11805013, 12075257 and the Fundamental Research Funds for the Central Universities under Grant No. 2018NTST09.

\bibliographystyle{apsrev}
\bibliography{reference}

\end{document}